\documentclass[aps,prd,amsmath,amssymb,twocolumn,preprintnumbers,nofootinbib]{revtex4}
\pdfoutput=1
\usepackage{graphicx}
\graphicspath{{./plots/}}
\usepackage{color}
\usepackage{hyperref}					
\usepackage{mathtools}					
\usepackage{slashed}					
\renewcommand{\epsilon}{\varepsilon}		

\newcommand{\beq}{\begin{eqnarray}}
\newcommand{\eeq}{\end{eqnarray}}

\newcommand{\bmp}{\noindent\begin{minipage}{16cm}}
\newcommand{\emp}{\end{minipage}\vskip 7mm} 

\renewcommand{\vec}{\mathbf} 
\DeclarePairedDelimiter{\ev}{.}{\vert} 	
\DeclareMathOperator{\tr}{Tr}				
\DeclareMathOperator{\re}{Re}				

\newcommand{\abs}[1]{\lvert #1 \rvert}
\def\simge{\mathrel{%
   \rlap{\raise 0.511ex \hbox{$>$}}{\lower 0.511ex \hbox{$\sim$}}}}
\def\simle{\mathrel{
   \rlap{\raise 0.511ex \hbox{$<$}}{\lower 0.511ex \hbox{$\sim$}}}}

\usepackage{dcolumn}
\usepackage{bm}
\usepackage{bbm}
\usepackage{subfigure}
\usepackage{pxfonts}

\definecolor{rossoCP3}{cmyk}{0,.88,.77,.40}

\def\lsim{\mathrel{\rlap{\lower4pt\hbox{\hskip1pt$\sim$}}
    \raise1pt\hbox{$<$}}}                
\def\gsim{\mathrel{\rlap{\lower4pt\hbox{\hskip1pt$\sim$}}
    \raise1pt\hbox{$>$}}}                

\newcommand{\be}{\begin{eqnarray}}
\newcommand{\ee}{\end{eqnarray}}

\newcommand{\halv}{\frac{1}{2}}  

\begin{document}
\includegraphics[width=3.cm]{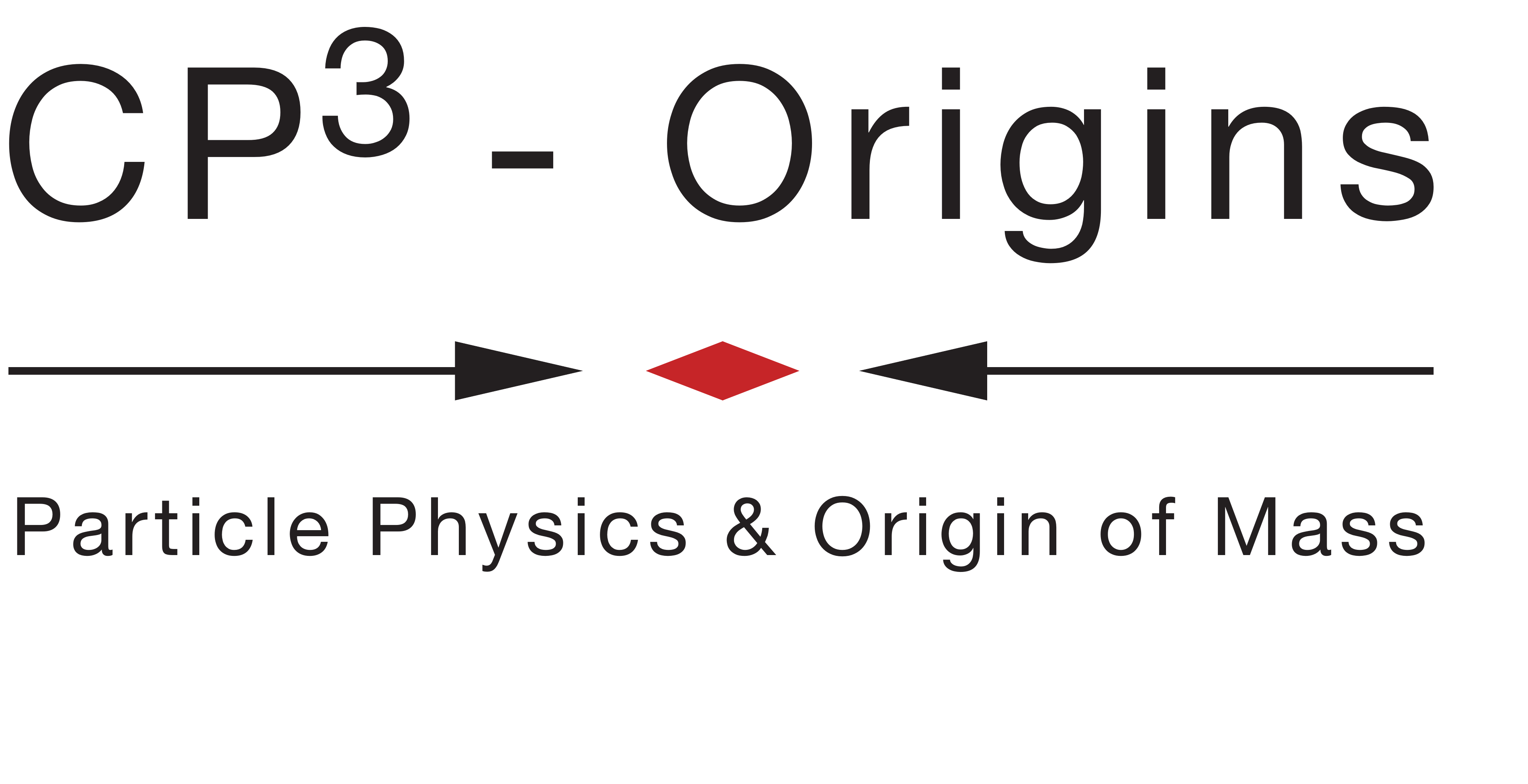} 
\title{\Large  \color{rossoCP3} S-parameter at Non-Zero Temperature and Chemical Potential}
\author{Ulrik Ish\o j {\sc S\o ndergaard}$^{\color{rossoCP3}{\varheartsuit}}$}
\email{sondergaard@cp3.sdu.dk} 
\author{Claudio {\sc Pica}$^{\color{rossoCP3}{\varheartsuit}}$}
\email{pica@cp3-origins.net} 
\author{Francesco {\sc Sannino}$^{\color{rossoCP3}{\varheartsuit}}$}
\email{sannino@cp3-origins.net} 
\affiliation{
$^{\color{rossoCP3}{\varheartsuit}}${ CP}$^{ \bf 3}${-Origins}, 
Campusvej 55, DK-5230 Odense M, Denmark.}
\begin{abstract}

We compute the finite-temperature and matter density corrections to the S-parameter at the one loop level. At non-zero temperature $T$ and matter density $\mu$ Lorentz symmetry breaks and therefore we suggest a suitable generalization of the S-parameter. 
By computing the plasma correction, we discover a reduction of the S-parameter in the physically relevant region of small external momenta for any non-zero $\mu$ and $T$. In particular, the S-parameter vanishes at small $m/T$, where $m$ is the mass of the fermions, due to the finite extent of the temporal direction.
Our results are directly applicable to the determination of the S-parameter via first principle lattice simulations performed with anti-periodic boundary conditions in the temporal direction.
\\[.1cm]
{\footnotesize  \it Preprint: CP$^3$-Origins-2011-10}
\end{abstract}

\maketitle

\section{Introduction}
Models of dynamical electroweak symmetry breaking are quickly gaining momentum. The reason why these models featuring uncolored techniquarks are harder to constrain, even at the LHC, is that they typically predict a heavier spectrum, compared to the electroweak scale, of new particles\footnote{Light states can also be present, depending on the specific dynamics. Here we are referring to the QCD-like states.}. For this reason it is interesting to analyze the indirect constraints from electroweak precision data established at LEP. It is well known that any extension of the Standard Model replacing the electroweak symmetry breaking sector can be constrained analyzing the electroweak gauge bosons vacuum polarizations as proved long time ago by Kennedy and Lynn \cite{Kennedy:1990ib}. Indirect constraints were therefore set using LEP results via the S, T and U parameters, or any linear combinations of them in \cite{Peskin:1990zt,Peskin:1991sw,Kennedy:1990ib,Altarelli:1990zd}.
It is therefore crucial to gain as much information as possible about these important correlators. In particular, the S-parameter is especially relevant for models of electroweak symmetry breaking. Several estimates have appeared in the literature making use of model computations \cite{Appelquist:1998xf,Kurachi:2006mu,Kurachi:2006ej,Sundrum:1991rf,Hong:2006si,Hirn:2006nt,Agashe:2007mc,Sannino:2010ca,Sannino:2010fh,DiChiara:2010xb,Round:2011nh,Dietrich:2008up,Anguelova:2011bc} and/or first-principle lattice computations \cite{Shintani:2008qe,Boyle:2009xi}. The latter ones are necessarily carried out on finite lattices, i.e. at finite volume and temperature. Surprisingly the finite size corrections to such parameters have not yet been investigated. Here we analyze the impact of the finite-temperature corrections on the S-parameter. 

This correlator can be determined for any asymptotically free gauge theory with matter transforming according to a given representation of a generic underlying gauge group. Once the number of colors of the new gauge dynamics is fixed, we can have a number of distinct phases in which the theory can exist. {}For example, it can display large distance conformality or break the global chiral symmetries spontaneously. In both cases this correlator is well defined \cite{Sannino:2010ca,Sannino:2010fh,DiChiara:2010xb} and worth computing. 

 We provide here the first one loop determination of the S-parameter at non-zero matter density $\mu$ and temperature $T$.  These computations are either under perturbative control when applied to the upper end of the conformal window or can be viewed as naive estimates, \`a la Peskin and Takeuchi,  when used below the conformal window. Since at non-zero temperature and matter density Lorentz symmetry breaks we suggest a suitable generalization of the S-parameter and investigate the various limits in the ratios of the relevant energy scales. 

In Sect.~\ref{sect2} we compute the plasma corrections to the S-parameter and we show that, at small external momenta, it is reduced with respect to the zero $T$ and $\mu $ case. We consider different limits for which analytical expressions can be derived, better elucidating the physical results.

In Sect.~\ref{sect3} we discuss the relevance of our results for lattice determinations of the S-parameter and we finally conclude in Sect.~\ref{sect4}.

A detailed derivation of our results is provided in the Appendices.

\section{The S-parameter at non-zero temperature and chemical potential\label{sect2}}
The definition of the S-parameter used by Polonsky and Su \cite{He:2001tp} and in \cite{Sannino:2010ca,DiChiara:2010xb} is
\be
S=-16 \pi \frac{\Pi_{3Y} (k^2)-\Pi_{3Y} (0)}{k^2}\ ,\label{S_parameter_org_def}
\ee
where $\Pi_{3Y}$ is the vacuum polarization of the thrid component of the isospin into the hypercharge current, and we have used, as a reference point, an external momentum $k^2$, instead of the usual Z boson mass. The S-paramter in Eq.~\eqref{S_parameter_org_def} also depends on the up- and down-type fermions masses $m_u$, $m_d$.

The definition by Peskin and Takeuchi \cite{Peskin:1990zt} is recovered in the $k^2\rightarrow 0$ limit:
\be
S=-16 \pi \ev*{\frac{d\Pi_{3Y} (k^2)}{dk^2}}_{k^2=0}\ .\label{S_parameter_PT}
\ee
At non-zero temperature, Lorentz invariance breaks, and therefore one should differentiate between the temporal and spatial components of the 4-momentum $k$.  Also several new scales will enter in the definition of the S-parameter.

The partition function of a four dimensional field theory with (anti)periodic boundary conditions in the euclidianized temporal direction coincides with the partition function of a three dimensional theory at finite temperature. The compactification of the temporal direction makes $k^0$ discrete. For a pedagogical introduction of the formalism we refer to the book by J. I. Kapusta  and C.~Gale~\cite{Kapusta:2006pm}. For a system in thermal equilibrium at temperature $T$ and chemical potential $\mu$, the allowed frequencies have values $k^0=i\omega_n +\mu$ where $\omega_n=2n\pi T$ for bosons and $\omega_n=(2n+1)\pi T$ for fermions.

At non-zero temperature, we extend the definition of the Polonsky-Su S-parameter in the following way:
\small\begin{align}
&S^{\mu\nu}(m_u,m_d,x,k^2,T,\mu)=\nonumber \\
&-16 \pi \frac{\Pi_{3Y}^{\mu\nu} (m_u,m_d,x, k^2,T,\mu)-\Pi_{3Y}^{\mu\nu} (m_u,m_d,x,0,T,\mu)}{{k}^2},\label{S_parameter_new_def}
\end{align}\normalsize
with the reference external 4-momentum given by $(k_0,\vec{k})$ with $k_0= x |\vec{k}|$ and $k^2=k_0^2-\vec{k}^2$.
Note that the external 4-momentum can also be written as
\begin{equation}
\begin{pmatrix}
k_0\\
| \vec k |
\end{pmatrix}=
\begin{pmatrix}
k \cosh \eta\\
k \sinh \eta
\end{pmatrix}\ ,\ x=\coth \eta \ge 1\ ,
\end{equation}
with $\eta$ the rapidity of the 4-vector $(k_0,\vec{k})$. As expected from the breaking of Lorentz invariance, the generalized S-parameter in Eq.~\eqref{S_parameter_new_def} does not depend only on $k^2$, but also on $x=\coth\eta$.

We also extend the Peskin-Takeuchi definition at finite temperature, by taking the limit $k^2\rightarrow 0$ of Eq.~\eqref{S_parameter_new_def}:
\be
S^{\mu\nu}(m_u,m_d,x,T,\mu)=-16 \pi \ev*{\frac{d\Pi^{\mu\nu}_{3Y} (m_u,m_d,x, k^2,T,\mu)}{d{k}^2}}_{k^2=0}.\label{S_parameter_PT_FT}
\ee

The complete expressions for the finite temperature $\Pi^{\mu\nu}_{3Y} (m_u,m_d,x, k^2,T,\mu)$ and the S-parameter are derived in Appendix \ref{App_derivation}. 

\begin{figure}[tb]
\begin{center}
\includegraphics[width=\columnwidth]{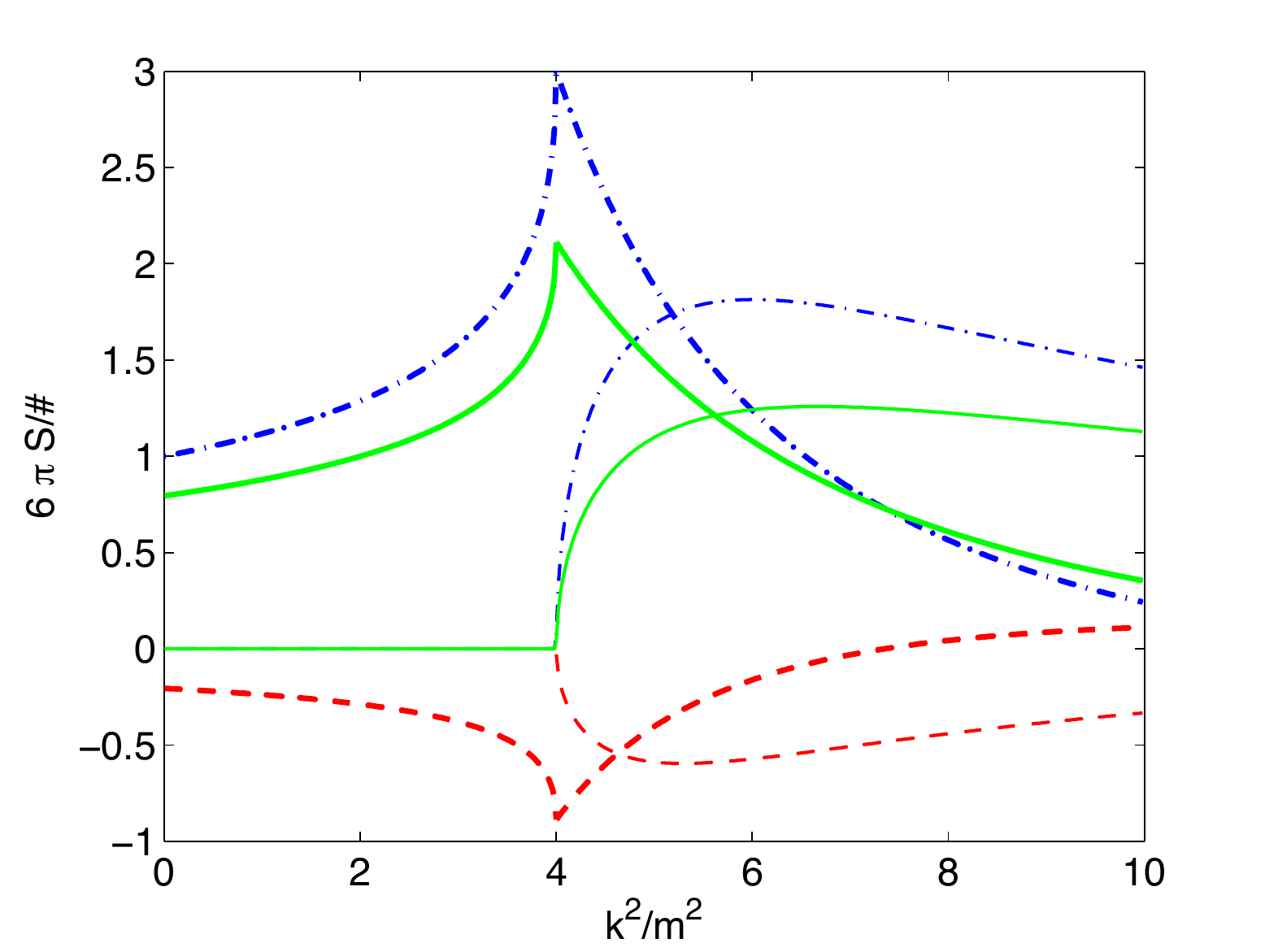}
\caption{The finite temperature S-parameter~\eqref{S_parameter_new_def} (real part: thick solid line, imaginary part: thin solid line) is given by the sum of the zero-temperature part $S_0$ (dashed-dotted lines) and a plasma contribution $S_+$~\eqref{Eq:S-plasma} (dashed lines). In this plot $\coth \eta =x=\sqrt{2}$, $\mu=0$ and $\beta m =1$.  
\label{Sx_Tinf}}
\end{center}
\end{figure}

The S-parameter being a pure number can only depend on adimentional ratios of the physical scales entering in its definition. 
In the case of degenerate u- and d-type fermions we have $m_u=m_d=m$, and the S-parameter will only depend on the pure numbers $x$, $k^2/m^2$, $\beta m$ and $\beta\mu$: $S(m,x,k^2,T,\mu)=S(x,k^2/m^2,\beta m,\beta\mu)$.
As shown in the Appendix \ref{App_derivation}, the total S-parameter is given by its zero-temperature and zero chemical potential part $S_0$ plus a plasma contribution $S_+$: $S = S_0 + S_+$. 
The explicit expression of the plasma contribution to the S-parameter is given by:
\begin{widetext}
\begin{align}
S_+(x,(k/m)^2,\beta m,\beta\mu)
=\frac{\sharp}{6\pi} \int_0^\infty dq q^2 24 \frac{\tilde F (\beta m, \beta \mu,q)}{\sqrt{1+q^2}}\left(
\frac{\sqrt{x^2-1}\ln \left( \frac{u_+}{u_-} \right)}{4q \frac{k}{m}} +   \frac{x^2-1}{2 \left(q^2 \left(x^2-1\right)+x^2\right)}  \right)\left( \frac{m}{k} \right)^2\ ,\label{Eq:S-plasma}
\end{align}
where we used the shorthand notation:
\begin{equation}
u_\pm = \pm 4 \frac{k}{m} q \sqrt{x^2-1} +\frac{k^2}{m^2}(x^2-1)-4\left( x^2+q^2 (x^2-1) \right)\ ,
\end{equation}
and the function $\tilde F$ is defined as:
\begin{equation}
\tilde F (\beta m, \beta \mu,q)=\frac{1}{\exp\left( \beta m \sqrt{1+q^2} -\beta \mu \right) +1}+\frac{1}{\exp\left( \beta m \sqrt{1+q^2} +\beta \mu \right) +1}\ .
\end{equation}
\end{widetext}
In Eq.~\eqref{Eq:S-plasma},  the factor $\sharp= d[r]N_f/2$ accounts for $N_D=N_f/2$ doublets of fermions in the representation $r$ with dimension $d[r]$ of the gauge group. 
To understand the plasma contribution to the S-parameter, we plot in Fig.\ref{Sx_Tinf} the finite temperature S-parameter, in the case $\mu=0$, as a function of $k^2/m^2$ for a particular value of $\beta m$. The zero-temperature part $S_0$ was computed in \cite{Sannino:2010ca} and it is displayed for comparison by the dashed-dotted lines (the real part corresponds to the thick line, while the imaginary part to the thin line). The plasma contribution $S_+$ is shown by the dashed lines, and the total S-parameter is diplayed by the solid lines. As we will show below with a direct computation, $S_+$ vanishes in the zero-temperature limit as expected. At non-zero temperature the structure of the real and imaginary parts of $S_+$ is similar to the one of $S_0$, and, at small $k^2/m^2$, reduces the total S-parameter. When increasing the temperature, $S_+$ grows in absolute value, and in the limit $\beta m\rightarrow 0$ exactly cancels the zero-temperature part, so that the S-parameter vanishes in this limit. 
In the case of zero chemical potential, this cancellation is encoded directly in the Matsubara frequency sum formula \eqref{Eq:matsubara}. 
Physically one can understand this cancellations by recalling that in this limit the fermions decouple.

Adding a non-zero chemical potential does not alter this picture, as we shall see below, and a similar reduction at small $k^2/m^2$ is observed.

\subsection{The $k^2/m^2\rightarrow 0$ limit}
Starting from Eq.~\eqref{Eq:S-plasma} we expand the term in parenthesis as a power series in $k^2/m^2$.
We have:
\begin{align}
\sqrt{x^2-1}\frac{\ln \left( \frac{u_+}{u_-} \right)}{4q \frac{k}{m}}+\frac{x^2-1}{2 \left(q^2 \left(x^2-1\right)+x^2\right)}=\sum_{i=1}^\infty a_i \left(\frac{k^2}{m^2}\right)^i \ ,\label{S_exp}
\end{align}
where the first coefficients in the expansion are
\begin{align}
a_1&=-\frac{\left(x^2-1\right)^2 \left(q^2 \left(3 x^2+1\right)+3 x^2\right)}{24 \left(q^2
   \left(x^2-1\right)+x^2\right)^3}, & \\
   a_2&=-\frac{\left(x^2-1\right)^3 \left(q^4 \left(5 x^4+10 x^2+1\right)+10
   q^2 \left(x^4+x^2\right)+5 x^4\right)}{160 \left(q^2
   \left(x^2-1\right)+x^2\right)^5} \ .
\end{align}
By inserting the expansion in Eq.~\eqref{S_exp} into Eq.~\eqref{Eq:S-plasma}, we obtain
\begin{align}
S_+=\frac{\sharp}{6\pi} \sum_{j=0}^\infty C_j  \left(\frac{k^2}{m^2}\right)^j \ ,
\end{align}
where
\begin{align}
C_j(x, \beta m, \beta \mu) = 24  \int_0^\infty dq q^2 \frac{\tilde F (\beta m, \beta \mu,q)}{\sqrt{1+q^2}} a_{j+1}.
\end{align}

The generalized S-parameter in Eq.~\eqref{S_parameter_PT_FT} can be easily read from the above expression:
\begin{align}
S(k^2/m^2\rightarrow 0)=\frac{\sharp}{6\pi}\left[ 1 +  C_0(x,\beta m,\beta \mu) \right] \ .\label{S_PT}
\end{align}

\begin{figure}[tb]
\begin{center}
\includegraphics[width=\columnwidth]{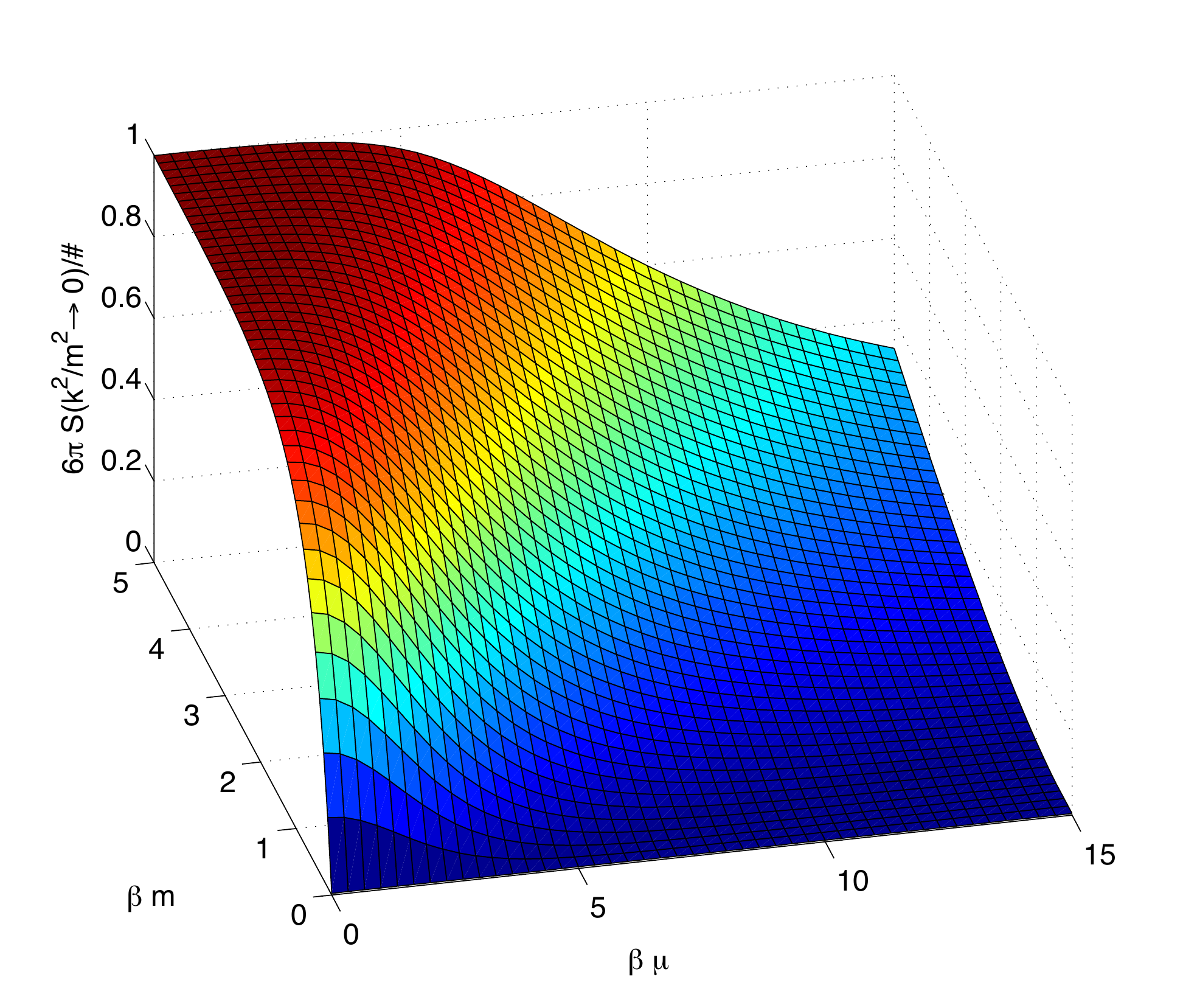}
\caption{The generalized S-parameter in Eq.~\eqref{S_parameter_PT_FT} at finite temperature and chemical potential for $\coth \eta=x=\sqrt{2}$.\label{Fig:S_PT}}
\end{center}
\end{figure}
We plot in Fig.~\ref{Fig:S_PT} the generalized S-parameter in Eq.~\eqref{S_parameter_PT_FT} and \eqref{S_PT} as a function of the temperature and chemical potential.
We find the interesting result that the S-parameter reduces in the presence of non-zero temperature and/or chemical potential. 

We have already discussed why the S-parameter is reduced at small $k^2/m^2$ at high temperature. The decrease as a function of the chemical potential is due to the fact that only one species of fermions survives at large $\mu$, effectively restoring chiral symmetry. At large chemical potential there will be instabilities at the Fermi surface leading to the breaking of the vacuum, which we are not considering here.

\subsection{High-temperature limit}
In the limit of large temperature $\beta m \rightarrow 0$, we expand $\tilde F$ as a power series in $\beta m$:
\begin{align}
 \tilde F(\beta m, \beta \mu) &= 1-\operatorname{sech}^2\left(\tfrac{1}{2}\beta \mu\right)\frac{1}{2} \sqrt{1+q^2} \beta m + \mathcal O(\beta^3 m^3) \ ,\label{Eq:Ftilde}
\end{align}
and insert this expansion in the integral defining $S_+$~\eqref{Eq:S-plasma}.
The first term in the expansion~\eqref{Eq:Ftilde}, constant in $\beta m$, gives a contribution which is identical in size but opposite in sign to the zero temperature S-parameter $S_0$. The second term, linear in $\beta m$, can also be computed leading to:  
\begin{equation}
S_+=-S_0+\operatorname{sech}^2\left(\tfrac{1}{2}\beta \mu\right)  \cosh (\eta)\, S_+^{(1)}(k^2/m^2)\left(\beta m\right)+\mathcal O(\beta^3 m^3)\ ,\label{Eq:S_HT}
\end{equation} 
where:
\begin{equation}
S_+^{(1)}(k^2/m^2) = \frac{3 \pi}{k^2/m^2}\left( 1 + i \sqrt{\frac{1}{4}\frac{k^2}{m^2}-1}\right)\ .
\end{equation}

In the limit $k^2/m^2\rightarrow 0$ the generalized S-parameter in Eq.~\eqref{S_parameter_PT_FT} is given by:
\begin{align}
S(k^2/m^2\rightarrow 0)=\frac{\sharp}{16}\operatorname{sech}^2\left(\tfrac{1}{2}\beta \mu\right) \cosh(\eta) \left(\beta m\right)  + \mathcal O(\beta^3 m^3)\ .
\end{align}

Interestingly we note that, in this limit, the contribution to S-parameter depending on $\beta\mu$, $\eta$, $k^2/m^2$ and $\beta m$ all factorize.

\subsection{Low-temperature limit}
When the ratio $\beta m$ is large enough, corresponding to the low-temperature limit of the theory, we expect to recover the familiar zero-temperature result. To show that this is indeed the case, we start from the expression for $S_+$ in Eq.\eqref{Eq:S-plasma} and perform a change of variable and rewrite the integral in terms of the new variable $Q=\beta m (\sqrt{1+q^2} - 1)$. We then expand the integrand as an asymptotic series for large $\beta m$.
In the final expression the dependence on the chemical potential factorizes, as in the case of the high-temperature expansion. This can be easily seen as, at large $\beta m$, the function $\tilde F$ appearing in  Eq.\eqref{Eq:S-plasma} can be approximated by:
 \begin{align}
 \tilde F(\beta m, \beta \mu) &\approx 2\cosh \left( \beta \mu\right)e^{-\beta m \sqrt{1+q^2}} \ .
 \end{align}

Keeping only the first term in the expansion of $S$ we obtain:
\begin{align}
S_+ = -\frac{\sharp}{6\pi} \frac{ \cosh (\beta \mu) \operatorname{sech}^{4}(\eta)}{1-\left(\frac{k/m}{2\cosh \eta}\right)^2} \frac{3\sqrt{2\pi}}{(\beta m)^{3/2}}\, e^{-\beta m} \cdot\left( 1+ \mathcal O \left( \frac1{\beta m}\right)\right)\ .\label{Eq:S_LT}
\end{align}
Higher order terms in the expansion can easily be obtained in the same way.

The final expression~\eqref{Eq:S_HT} contains an exponential Boltzmann-like suppression factor as expected. 

\begin{figure}[tb]
\begin{center}
\includegraphics[width=\columnwidth]{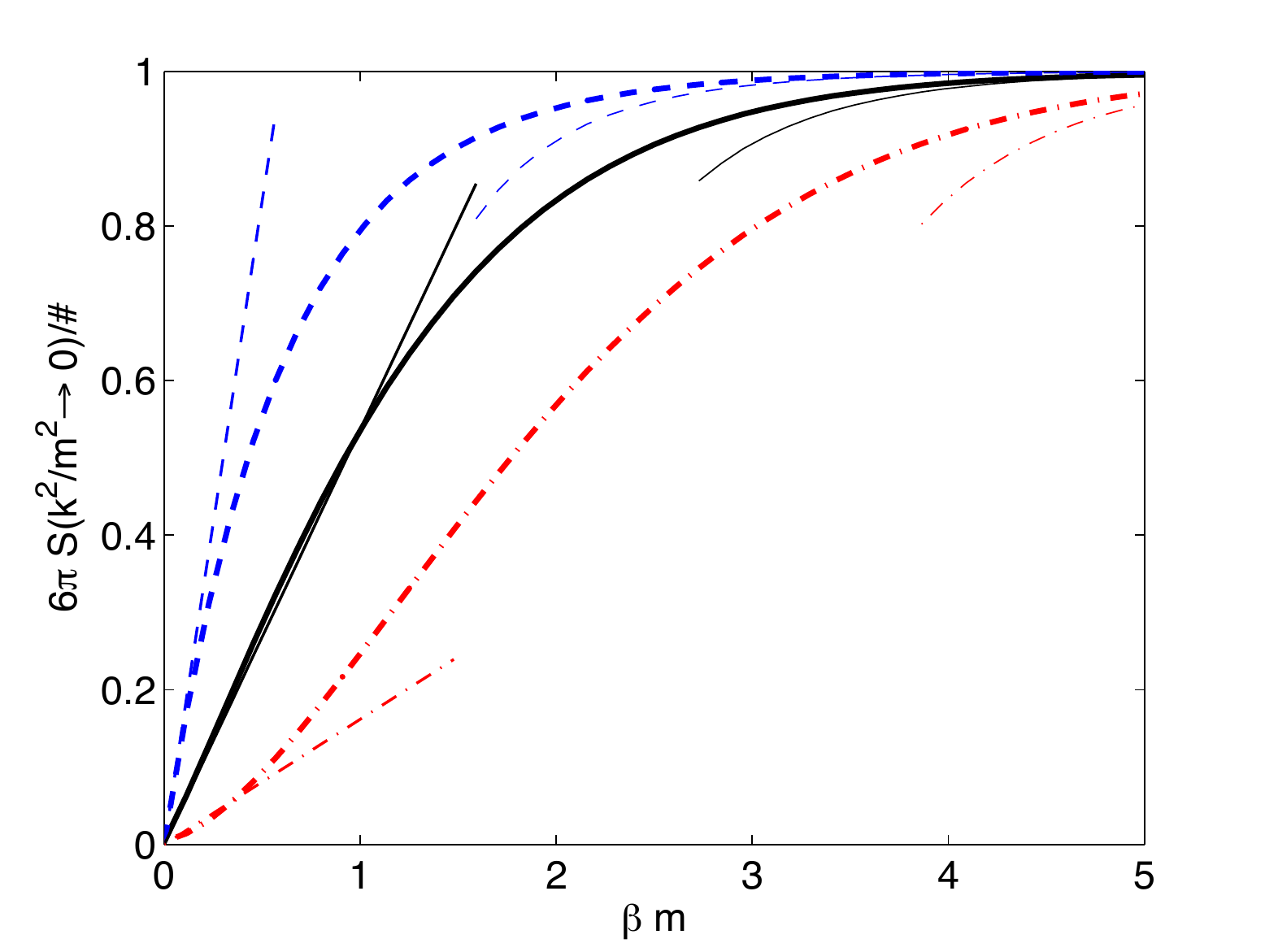}
\caption{The generalized S-parameter in Eq.~\eqref{S_parameter_PT_FT} (thick lines) at finite temperature and chemical potential as compared to the high-temperature~\eqref{Eq:S_HT} and low-temperature~\eqref{Eq:S_LT} expansions. The three different curves correspond to $\beta\mu = 0$ (top, dashed line), $\beta\mu=2.33$ (middle, solid line) and $\beta\mu=4.33$ (bottom, dashed-dotted line). In this plot we use $\coth\eta=\sqrt{2}$.\label{S_PT_APPR}}
\end{center}
\end{figure}

We compare the high- and low-temperature expansions in Eq.~\eqref{Eq:S_HT} and \eqref{Eq:S_LT} to the complete numerical result for the generalized S-parameter in the $k^2/m^2\rightarrow 0$ limit in Figure \ref{S_PT_APPR}.
The expansions converge within their respective domain of validity temperature regions.

\begin{figure}[tb]
\begin{center}
\includegraphics[width=\columnwidth]{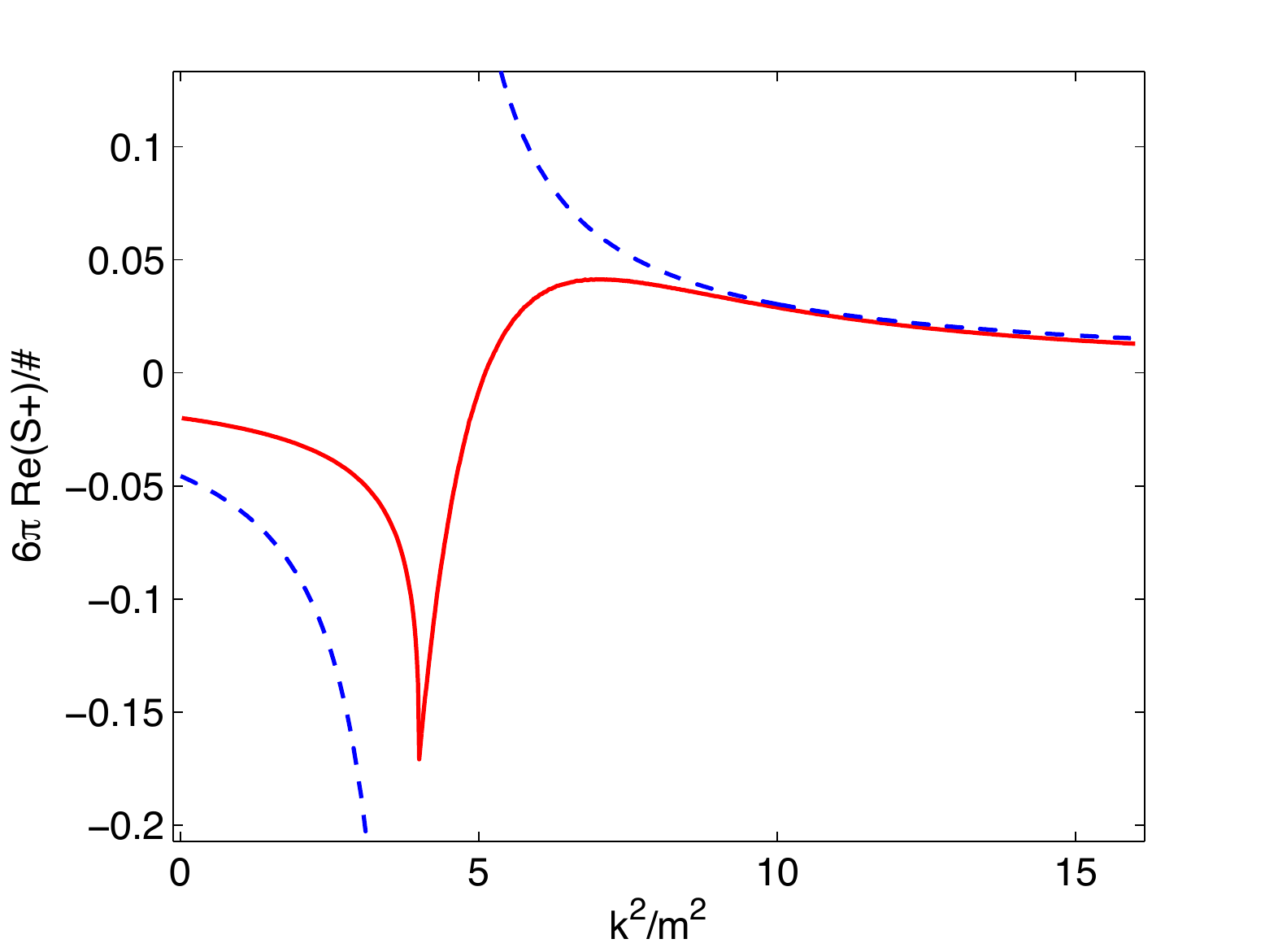}
\caption{The plasma contribution to the S-parameter~\eqref{Eq:S-plasma} (solid line) at finite temperature $\beta m =5$ and finite chemical potential $\beta \mu =3$ compared to the low-temperature expansion~\eqref{Eq:S_LT} (dashed line).\label{S_LT_APPR}}
\end{center}
\end{figure}

We also note that the accuracy of the low-temperature expansion in Eq.~\eqref{Eq:S_LT} depends critically on the value of $k^2/m^2$ as the coefficient of the first term in the expansion diverges for $k/m = 2 \cosh\eta$. 
This divergence is unphysical, as it is clear from the fact that it is not present in the full expression.
We compare in Fig.~\ref{S_LT_APPR} the value of $S_+$ with its asymptotic expansion at low-temperature for a finite, low value of the temperature $\beta m =5$ and at finite density $\beta \mu =3$. 
The deep at $k^2/m^2=4$ of $S_+$ is the kinematical threshold.

\begin{figure}[tb]
\begin{center}
\includegraphics[width=\columnwidth]{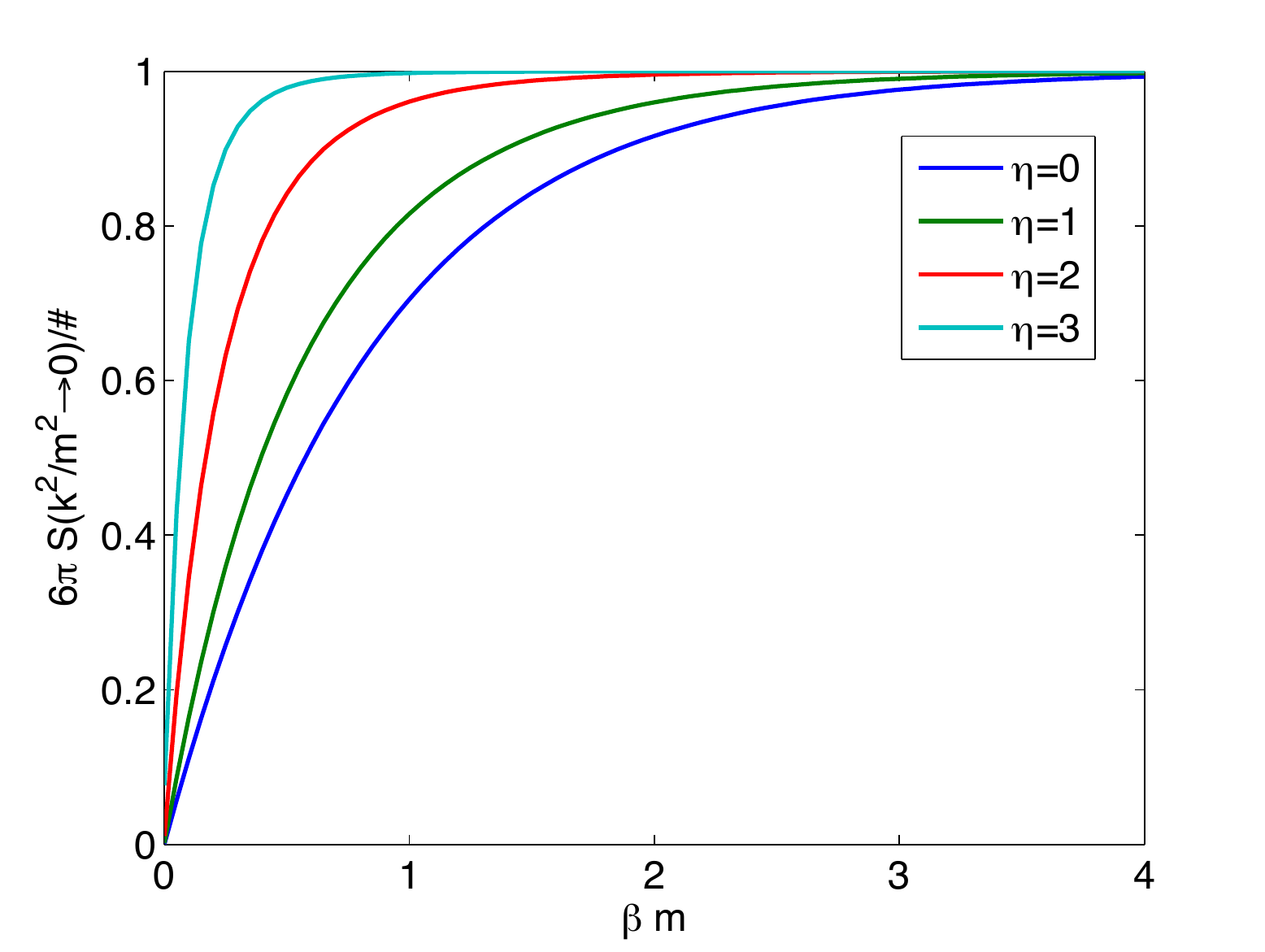}
\caption{The rapidity dependence of the S-parameter in the limit $k^2/m^2\rightarrow 0$ for $\mu =0$. The finite-temperature corrections vanish at large rapidity.\label{S_X_DEP1}}
\end{center}
\end{figure}

\begin{figure}[tb]
\begin{center}
\includegraphics[width=\columnwidth]{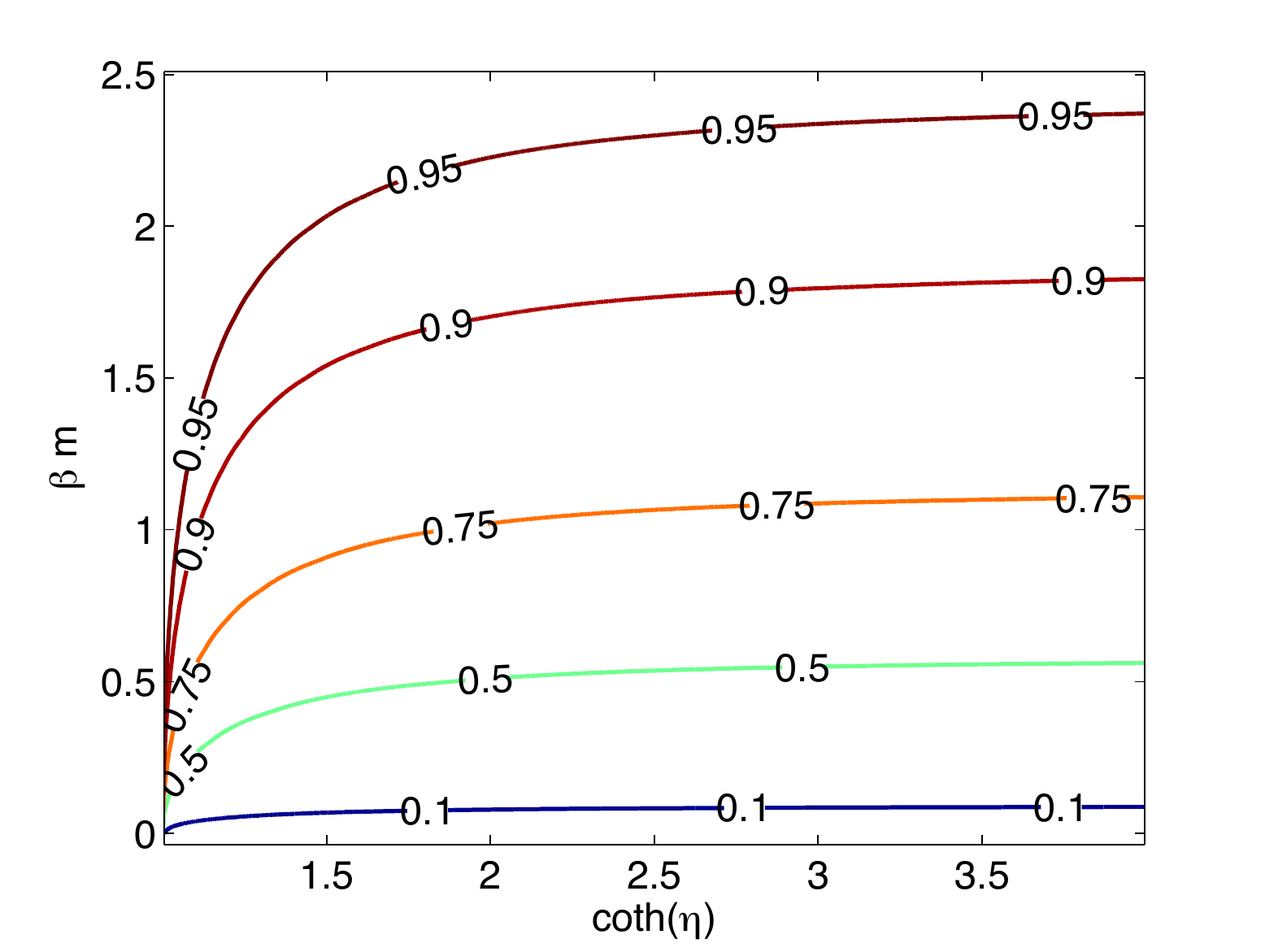}
\caption{Contour plot of $S/S_0$ at $\mu=0$.\label{S_X_DEP2}}
\end{center}
\end{figure}

\begin{figure}[tb]
\begin{center}
\includegraphics[width=\columnwidth]{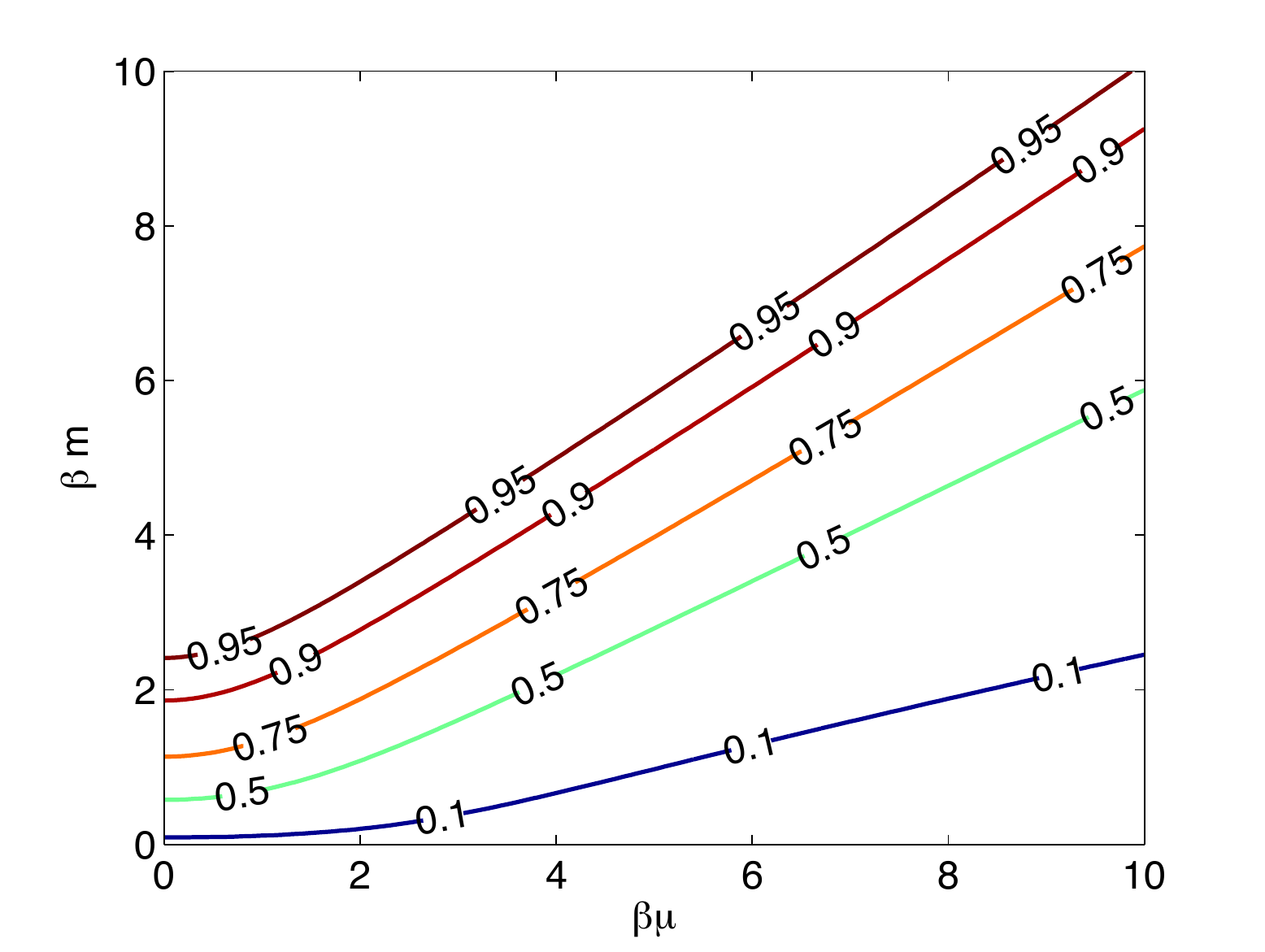}
\caption{Contour plot of $S/S_0$ for $\coth \eta = \sqrt{2}$ corresponding to Fig.~\ref{Fig:S_PT}.\label{S_X_DEP3}}
\end{center}
\end{figure}

\begin{figure}[tb]
\begin{center}
\includegraphics[width=\columnwidth]{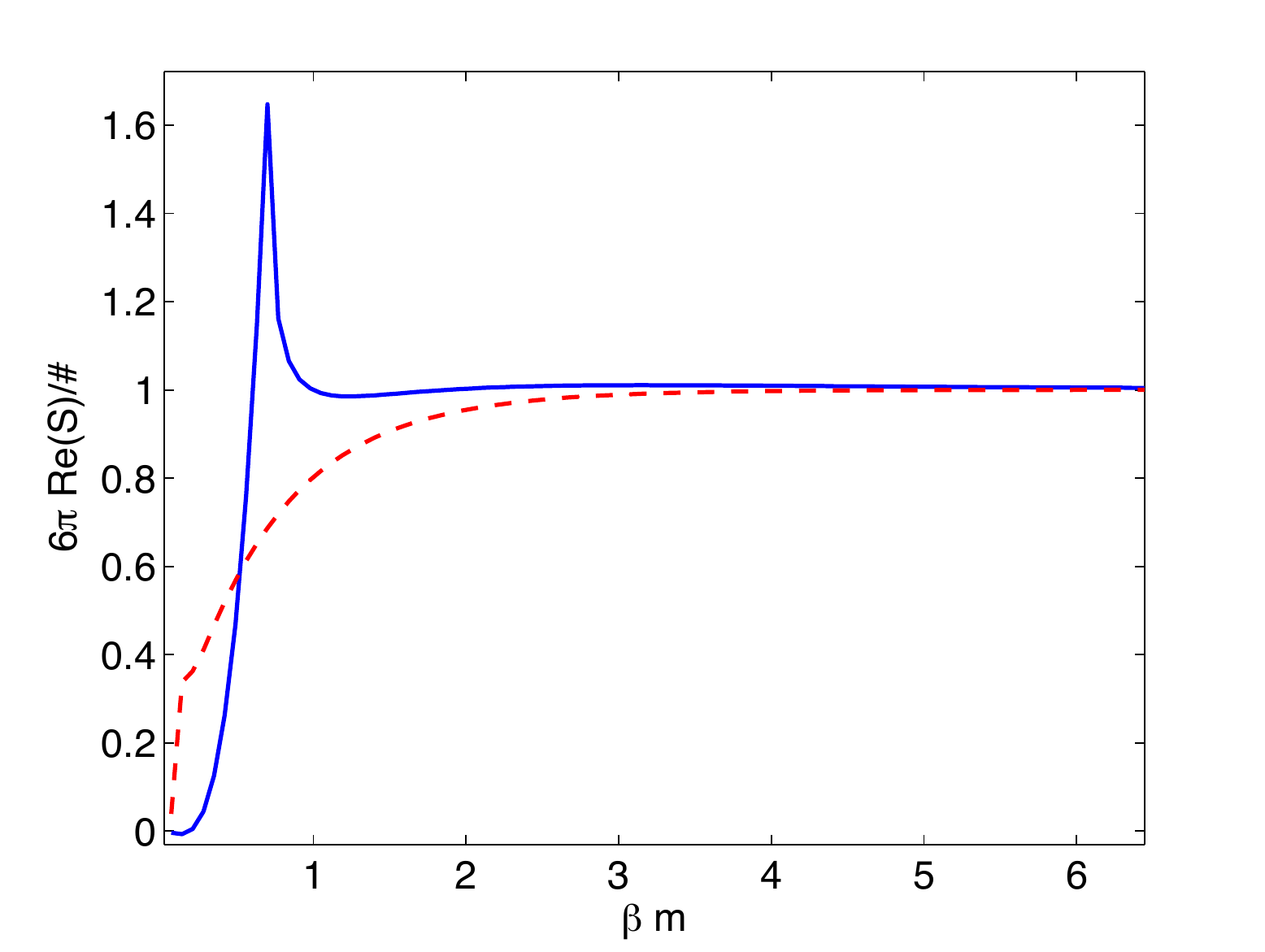}
\caption{S-parameter at finite temperature as a function of the fermions mass for $k^2\beta^2=0.1$ (dashed line) and $k^2\beta^2=2$ (solid line). In this figure $\coth\eta= \sqrt{2}$.\label{S_M_DEP}}
\end{center}
\end{figure}
\section{Applications to Lattice Field Theory\label{sect3}}

Our result can be used to estimate the finite-temperature corrections to the S-parameter as computed using numerical lattice simulations. These are typically performed at non-zero temperature, when anti-periodic boundary conditions are used in the temporal direction, so that it is crucial to disentagle the finite temperature effects to extrapolate to zero temperature. We demonstrated that the S-parameter in the phenomenologically relevant parameter region of small external momenta, is reduced by the effects of non-zero temperature. The smaller is $\beta m$ the larger the negative corrections. 
There is also a dependence on the rapidity $\eta$ which has to be taken into account. At large $\beta m$ such a dependence disappears, while it increases at smaller $\beta m$. This dependence is shown in Fig.~\ref{S_X_DEP1}.
In the large rapidity regime, corresponding to $x\rightarrow 1$, one approaches the zero temperature result. This can be better understood considering that in this limit the temporal and spacial momentum both diverge in order to keep $k^2$ finite, as required by our definition of the S-parameter.

To better elucidate the size of the corrections to the S-parameter, we plot in Fig.~\ref{S_X_DEP2} the contour lines of $S/S_0$ in the $\beta m - \coth\eta$ plane. We also show in Fig.~\ref{S_X_DEP3} the contour plots of $S/S_0$ as a function of $\beta m$ and $\beta \mu$, for $\coth \eta = \sqrt{2}$ corresponding to Fig.~\ref{Fig:S_PT}, 

To compare to the lattice results, we show in Fig.~\ref{S_M_DEP} the dependence on the fermion masses of the S-parameter at zero chemical potential and finite $T$ for two different values of the reference momentum $k^2$. The decrease in S at small $m$ is not due to a chiral restoration but the finite temperature corrections. From Fig.~\ref{S_M_DEP} it is possible to estimate the range of masses for which a reliable estimate of the S-parameter can be extracted from lattice results, assuming that the infinite volume regime has been reached.

\section{Conclusions\label{sect4}}
In this work we provided a suitable generalization of the S-parameter at non-zero temperature and chemical potential.
By computing the plasma contributions $S_+$ we discovered a reduction of the S-parameter in the physically relevant region of small $k^2/m^2$ for any non-zero $\mu$ and $T$.
Our results are directly applicable to the determination for the S-parameter via first principle lattice simulations performed with anti-periodic boundary conditions in the temporal direction.
In this case we find that the vanishing of the S-parameter at small $\beta m$ is due to the finite extent of the temporal direction.

\acknowledgments
The authors acknowledge the useful discussions with Tuomas Hapola and Stefano Di Chiara in the initial stages of this work.

\appendix
\section{Vacuum contribution to the S-parameter}
At zero temperature $T=0$ and chemical potential $\mu=0$, the perturbative expression for the S-parameter, as defined by Polonsky and Su, has been calculated at 1-loop in \cite{He:2001tp} and at 2-loop in \cite{Djouadi:1993ss}. At 1-loop order the Polonsky-Su S-parameter, as defined in Eq.~\eqref{S_parameter_org_def}, reads
\begin{multline}
S= \frac{N_D d[r]}{6\pi}\left\{2(4Y+3)z_u+2(-4Y+3)z_d-2Y\log\left(\frac{z_u}{z_d}\right)+\right.\\
+\left.\left[\left(\frac{3}{2}+2Y\right)z_u+Y\right] G(z_u)  
+\left[\left(\frac{3}{2}-2Y\right)z_d-Y\right] G(z_d) \right\}\,, \label{SatT0}
\end{multline}
with 
\begin{equation}
G(z)=-4\sqrt{4z-1}\,\arctan\left(\frac{1}{\sqrt{4z-1}}\right)\nonumber
\,,
\label{eq:Gfun}
\end{equation}
 where $Y$ is the hypercharge, $z_i=m_i^2/k^2$, $i=u,d$, $m_i$ is the mass of the fermionic species and $N_D=N_f/2$ is the number of doublets transforming under a technicolor group representation $r$ with dimension $d[r]$.

 \section{Detailed calculation of plasma contribution to the S-parameter}\label{App_derivation}
The S-parameter can be written in terms of the following vacuum polarization amplitudes:
\begin{center}
\begin{tabular}{rl}
\raisebox{9mm}{$\Pi_{LL}^{\mu\nu}(m_u,m_d)=$ }&%
\includegraphics[trim = 0 2.8cm 0mm 0mm, clip,width=4cm]{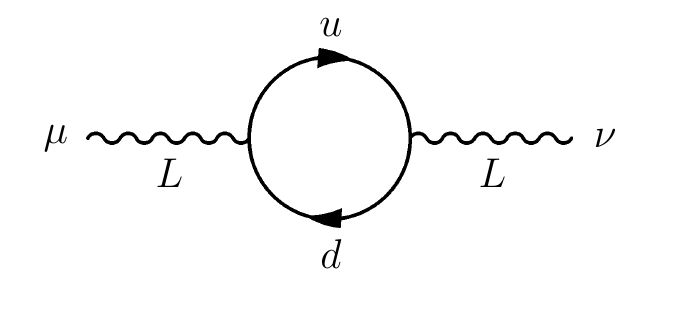}\raisebox{9mm}{\ ,}%
\\
\raisebox{8mm}{$\Pi_{LR}^{\mu\nu}(m_u,m_d)=$ }&%
\includegraphics[trim = 0  0  0 2.8cm, clip,width=4cm]{1loop.pdf}\raisebox{8mm}{\ ,}%
\end{tabular}
\end{center}
where the wavy lines represent electroweak gauge bosons, once we specify the hypercharge and isospin of the fermion doublets. We use the choice of table \ref{Assignments}.
\begin{table}[t]
\begin{center}
\begin{tabular}{|c|c|c|}
\hline
$\,$ & $T^3$ & $Y$ \\
\hline\hline
$u_L$ & $1/2$ & $y/2$ \\
\hline
$u_R$ & $0$ & $(y+1)/2$ \\
\hline
$d_L$ & $-1/2$ & $y/2$ \\
\hline
$d_R$ & $0$ & $(y-1)/2$ \\
\hline
\end{tabular}
\caption{Hypercharge and isospin assignments for the fermions. $y$ is an arbitrary real number.}
\end{center}
\label{Assignments}
\end{table}
The amplitude $\Pi_{3Y}^{\mu\nu}$, entering in the S-parameter, can then be constructed in terms of $\Pi_{LL}^{\mu\nu}$ and $\Pi_{LR}^{\mu\nu}$. We have:
\begin{align}
4\Pi_{3Y}^{\mu\nu}(k,m_u,m_d)
=&  y\big[ \Pi^{\mu\nu}_{LL}(k,m_u,m_u)+ \Pi^{\mu\nu}_{LR}(k,m_u,m_u)\big] \nonumber \\
&-y\big[ \Pi^{\mu\nu}_{LL}(k,m_d,m_d)+ \Pi^{\mu\nu}_{LR}(k,m_d,m_d)\big]\nonumber \\
&+ \Pi^{\mu\nu}_{LR}(k,m_u,m_u)+ \Pi^{\mu\nu}_{LR}(k,m_d,m_d)\ .\label{3Y_gen}
\end{align}
In the equations above we have suppressed the explicit dependence on the temperature $T$ and the chemical potential $\mu$. In the case of degenerate fermion masses, $\Pi_{3Y}^{\mu\nu}(k)$ becomes independent of the particular choice of hypercharge, and we have:
\be
\Pi_{3Y}^{\mu\nu}(k,m)= \halv\Pi^{\mu\nu}_{LR}(k,m)\ .
\ee
%
 \subsection*{Evaluation of diagrams}
The explicit expression for the amplitudes, at finite temperature and chemical potential, have the form:
\small\begin{align*}
\Pi_{LH}^{\mu\nu}&= T \sum_{l=-\infty}^\infty \int \frac{d^3\vec p }{(2\pi)^3} \tr \left[
\gamma^\mu P_L \frac{\slashed p + m}{p^2-m^2} \gamma^\nu P_H \frac{\slashed k +\slashed p + m}{(k+p)^2-m^2} 
\right],
\end{align*}\normalsize
 where $p^0=i(2l+1)\pi T+\mu$ and $H=L,R$. Let $N^{\mu\nu}_{LH}$ denote the numerator of the expression in square brackets. By evaluating the traces we get:
 \begin{align*}
 N^{\mu\nu}_{LL}&= 2\left[ p^\mu (p+k)^\nu + p^\nu (p+k)^\mu - p\cdot(p+k)g^{\mu\nu} \right],\\
 N^{\mu\nu}_{LR}&= 2m^2g^{\mu\nu}.
\end{align*}
Introducing the shorthand
\begin{equation*}
b=(p^0+k^0)^2-\vec p^2 - \vec k^2 -m^2\ ,
\end{equation*}
we can rewrite the amplitude in a more compact form:
\begin{align*}
\Pi_{LH}^{\mu\nu}= T \sum_{l=-\infty}^\infty \int \frac{d\vec p^3}{(2\pi)^3}  \frac{1}{p^2 - m^2}\cdot\frac{N_{LH}^{\mu\nu}}{b-2\abs{\vec p}\abs{\vec k} \cos \theta}\ .
\end{align*}
We now use the Matsubara frequency sum formula (see e.g. \cite{Kapusta:2006pm})
\begin{align}
T\sum_{l=-\infty}^\infty f(p^0)=
\frac{i}{2\pi}& \int_{-i\infty +\mu+\epsilon}^{+i\infty +\mu+\epsilon} dp^0 f(p^0) \tilde f (p^0-\mu)\nonumber\\
&+\frac{i}{2\pi}\int_{-i\infty +\mu-\epsilon}^{+i\infty +\mu-\epsilon} dp^0 f(p^0) \tilde f (\mu-p^0) \nonumber\\
&-\frac{i}{2\pi} \sqint_C dp^0 f(p^0)-\frac{i}{2\pi} \int_{-i\infty}^{i\infty} dp^0 f(p^0)\ ,\label{Eq:matsubara}
\end{align}
to evaluate the sum over the frequencies. In this expression $C$ is a rectangular path $(i \infty, - i\infty + \mu)$ going counter clockwise and $\tilde f(p^0)=(e^{\beta p^0}+1)^{-1}$ where $\beta=T^{-1}$. This formalism allows for a neat separation of the vacuum contribution $\Pi_{LH,0}^{\mu\nu}=\ev*{\Pi_{LH}^{\mu\nu}}_{T,\mu =0}$ and the plasma contribution $\Pi_{LH,+}^{\mu\nu}=\Pi_{LH}^{\mu\nu}-\Pi_{LH,0}^{\mu\nu}$. The last term in the sum formula gives the zero temperature and chemical potential, and reads:
\be
\Pi_{LH,0}^{\mu\nu}(k)=-i\int \frac{d^4p}{(2\pi)^4}\frac{1}{p^2 - m^2}\cdot\frac{N_{LH}^{\mu\nu}}{b-2\abs{\vec p}\abs{\vec k} \cos \theta} .\nonumber
\ee
For evaluation of this term we refer to standard textbooks, see e.g. Peskin and Schroeder \cite{Peskin:1995ev}. The final result for this integral corresponds to the S-parameter as stated in \eqref{SatT0}.

In this paper we compute $\Pi_{LH,+}^{\mu\nu}$. We first consider the case where no poles are contained inside the closed path $C$. This depends on the values of the chemical potential $\mu$ and the external momentum.  In this case only the first two integrals contribute to the result:
\small\begin{align*}
\Pi_{LH,+}^{\mu\nu}=&i \int_{-i\infty +\mu+\epsilon}^{i\infty+\mu+\epsilon}\frac{dp^0}{2\pi} \int \frac{d^3\vec p}{(2\pi)^3} \frac{1 }{ p^2-m^2}\frac{N_{LH}^{\mu\nu}}{b-2\abs{\vec p}\abs{\vec k} \cos \theta} \tilde f(p^0-\mu)\\
&+
i \int_{-i\infty+\mu -\epsilon}^{i\infty+\mu-\epsilon}\frac{dp^0}{2\pi} \int \frac{d^3\vec p}{(2\pi)^3} \frac{1 }{ p^2-m^2}\frac{N_{LH}^{\mu\nu}}{b-2\abs{\vec p}\abs{\vec k} \cos \theta}  \tilde f(\mu-p^0).
\end{align*}\normalsize
Changing to spherical coordinates the angular part of the  integration over $\vec p$ can be performed (note that also $N_{LL}^{\mu\nu}$ depends on the angle $\theta$). The results for the two amplitudes are:
\small\begin{align*}
\Pi_{LL,+}^{\mu\nu}=&i\int_{-i\infty + \mu+\epsilon}^{i\infty + \mu +\epsilon} \frac{dp^0}{2 \pi}\int_0^{\infty}\frac{d\abs{\vec p} \abs{\vec p}^2}{(2\pi)^2} \frac{2 \tilde f (p^0-\mu)}{p^2-m^2}     \\  &\qquad \quad   \left[  \frac{a^{\mu\nu}+b g^{\mu\nu}}{4\abs{\vec p}\abs{\vec k}} \ln\left( \frac{ b+2 \abs{\vec p}\abs{\vec k}}{b-2\abs{\vec p}\abs{\vec k}}\right)  - g^{\mu\nu}\right] \\
&+i\int_{-i\infty + \mu-\epsilon}^{i\infty + \mu -\epsilon} \frac{dp^0}{2 \pi}\int_0^{\infty}\frac{d\abs{\vec p} \abs{\vec p}^2}{(2\pi)^2} \frac{2 \tilde f (\mu-p^0)}{p^2-m^2}     \\  &\qquad \quad   \left[  \frac{a^{\mu\nu}+b g^{\mu\nu}}{4\abs{\vec p}\abs{\vec k}} \ln\left( \frac{ b+2 \abs{\vec p}\abs{\vec k}}{b-2\abs{\vec p}\abs{\vec k}}\right)  - g^{\mu\nu}\right] ,
\end{align*}\normalsize
\small\begin{align*}
\Pi_{LR,+}^{\mu\nu}&=i\int_{-i\infty + \mu+\epsilon}^{i\infty + \mu +\epsilon} \frac{dp^0}{2 \pi}\int_0^{\infty}\frac{d\abs{\vec p} \abs{\vec p}^2}{(2\pi)^2} \frac{2 \tilde f (p^0-\mu)}{p^2-m^2}   
\frac{N_{LR}^{\mu\nu}}{4\abs{\vec p}\abs{\vec k}} \ln\left( \frac{ b+2 \abs{\vec p}\abs{\vec k}}{b-2\abs{\vec p}\abs{\vec k}}\right) \\
&+i\int_{-i\infty + \mu-\epsilon}^{i\infty + \mu -\epsilon} \frac{dp^0}{2 \pi}\int_0^{\infty}\frac{d\abs{\vec p} \abs{\vec p}^2}{(2\pi)^2} \frac{2 \tilde f (\mu-p^0)}{p^2-m^2}    
 \frac{N_{LR}^{\mu\nu}}{4\abs{\vec p}\abs{\vec k}} \ln\left( \frac{ b+2 \abs{\vec p}\abs{\vec k}}{b-2\abs{\vec p}\abs{\vec k}}\right)\ ,
\end{align*}\normalsize
where 
\small\begin{equation}
a^{\mu\nu}=2 p^\mu (p^\nu + k^\nu ) +2 p^\nu (p^\mu + k^\mu )- 2 p^0 (p^0+k^0)g^{\mu\nu}+2\abs{\vec p}^2 g^{\mu\nu}\ .\nonumber
\end{equation}\normalsize
Evaluating the integrals in $p^0$ by closing the contours at infinity gives:
\begin{align*}
\Pi_{LL,+}^{\mu\nu}&= -2\int_0^{\infty}\frac{d\abs{\vec p} \abs{\vec p}^2}{(2\pi)^2} \frac{ \tilde f (E_{\vec p}-\mu)+\tilde f (E_{\vec p}+\mu)}{E_{\vec p}}\\&
\re \left[  \frac{      a^{\mu\nu}+b g^{\mu\nu}     }{4\abs{\vec p}\abs{\vec k}} \ln\left( \frac{ b+2 \abs{\vec p}\abs{\vec k}}{b-2\abs{\vec p}\abs{\vec k}}\right)  - g^{\mu\nu}\right]_{p^0=E_{\vec p}} ,\\
\Pi_{LR,+}^{\mu\nu}&= -2\int_0^{\infty}\frac{d\abs{\vec p} \abs{\vec p}^2}{(2\pi)^2} \frac{ \tilde f (E_{\vec p}-\mu)+\tilde f (E_{\vec p}+\mu)}{E_{\vec p}}\\&
\re \left[  \frac{      N_{LR}^{\mu\nu}     }{4\abs{\vec p}\abs{\vec k}} \ln\left( \frac{ b+2 \abs{\vec p}\abs{\vec k}}{b-2\abs{\vec p}\abs{\vec k}}\right) \right]_{p^0=E_{\vec p}} .
\end{align*}
In the above expression we have:
\begin{align*}
\re f(k^0) &= \halv [ f(k^0)+f(-k^0) ]\ ,\\
E_{\vec p}&=\sqrt{\abs{\vec p}^2+m^2}\ .
\end{align*}
The computation was performed in the case where no poles reside in the interior of $C$. However it is straightforward to verify that this result holds also if some of the poles are located inside the contour $C$.

We are now ready to piece together the S-parameter from \eqref{3Y_gen} and the definition in Eq.s~\eqref{S_parameter_new_def}, \eqref{S_parameter_PT_FT}.
We first write the expression for $\Pi_{3Y,+}$: 
\begin{widetext}
\small\begin{align*}
\Pi_{3Y,+}^{\mu\nu}(k^0,\abs{\vec k},m_u,m_d)=\frac{y}{2}\re
\int_0^{\infty}\frac{d\abs{\vec p} \abs{\vec p}^2}{(2\pi)^2} 
\Bigg\{
&\frac{ \tilde f (E_{\vec p,d}-\mu)+\tilde f (E_{\vec p,d}+\mu)}{E_{\vec p,d}} \cdot
\left[  \frac{      a^{\mu\nu}+b_d g^{\mu\nu} +\left( 1-\frac{1}{y}\right) N_{LR,d}^{\mu\nu}    }{4\abs{\vec p}\abs{\vec k}}
 \ln\left( \frac{ b_d+2 \abs{\vec p}\abs{\vec k}}{b_d-2\abs{\vec p}\abs{\vec k}}\right)  - g^{\mu\nu}\right]_{p^0=E_{\vec p,d}}  \\
& -\frac{ \tilde f (E_{\vec p,u}-\mu)+\tilde f (E_{\vec p,u}+\mu)}{E_{\vec p,u}}\cdot
\left[  \frac{      a^{\mu\nu}+b_u g^{\mu\nu} +\left( 1+\frac{1}{y}\right)N_{LR,u}^{\mu\nu}    }{4\abs{\vec p}\abs{\vec k}}
 \ln\left( \frac{ b_u+2 \abs{\vec p}\abs{\vec k}}{b_u-2\abs{\vec p}\abs{\vec k}}\right)  - g^{\mu\nu}\right]_{p^0=E_{\vec p,u}} \Bigg\}\ ,
\end{align*}\normalsize
where the subscript $u,d$ denotes which mass is used. 
To write down the explicit expression for the S-parameter in Eq.~\eqref{S_parameter_new_def}, we now take the limit $k\rightarrow 0$ keeping $k^0=\cosh (\eta)k, \abs{\vec{k}}=\sinh (\eta) k$:
\small\begin{align*}
\lim_{k\rightarrow 0}\Pi_{3Y,+}^{\mu\nu}(\cosh (\eta)k,\sinh (\eta)k,m_u,m_d)=\frac{y}{4}
\int_0^{\infty}\frac{d\abs{\vec p} \abs{\vec p}^2}{(2\pi)^2} 
\Bigg\{
&
 \frac{ \tilde f (E_{\vec p,d}-\mu)+\tilde f (E_{\vec p,d}+\mu)}{E_{\vec p,d}} \cdot 
\left[  \frac{      a^{\mu\nu}+\left( 1-\frac{1}{y}\right) N_{LR,d}^{\mu\nu}    }{-2(m^2 \cosh^2(\eta) + p^2)}
  - g^{\mu\nu}\right]_{p^0=E_{\vec p,d}, k=0}  \\
&
-\frac{ \tilde f (E_{\vec p,u}-\mu)+\tilde f (E_{\vec p,u}+\mu)}{E_{\vec p,u}}\cdot
\left[  \frac{      a^{\mu\nu}+\left( 1+\frac{1}{y}\right)N_{LR,u}^{\mu\nu}    }{-2(m^2 \cosh^2(\eta)+ p^2)}
  - g^{\mu\nu}\right]_{p^0=E_{\vec p,u},k=0} \Bigg\}. 
\end{align*}\normalsize
\end{widetext}
The plasma contribution to the S-parameter $S_+$ can be computed from the explicit expressions for $\Pi_{3Y,+}$ and $\lim_{k\rightarrow 0}\Pi_{3Y,+}$ above
\begin{align*}
S_+(k^2,x,m_u,m_d)=-16 \pi \frac{\Pi_{3Y,+}-\lim_{k\rightarrow 0}\Pi_{3Y,+}}{k^2} \ ,
\end{align*}
where $x=\coth(\eta)$ and $\Pi_{3Y,+}$ is the transverse part of $\Pi_{3Y,+}^{\mu\nu}$.


\end{document}